\documentclass[aps,floatfix,twocolumn,final,prl,showpacs]{revtex4}

\usepackage{epsfig}
\usepackage{graphicx}
\usepackage[ansinew]{inputenc}

\begin{document}

\title{FFLO State and Peak Effect Dynamics in CeCoIn$_{5}$: Magnetization Studies.}

\author{X. Gratens $^{1}$, L. Mendonça Ferreira $^{2}$, Y. Kopelevich $^{2}$, N. F. Oliveira Jr.$^{1}$, P. G. Pagliuso $^{2}$,
R. Movshovich $^{3}$, R. R. Urbano $^{3}$, J. L. Sarrao $^{3}$, J. D. Thompson $^{3}$ } \affiliation{$^{1}$
Instituto de Física, Universidade de São Paulo, 05315-970, São Paulo, Brazil} \affiliation{$^{2}$Instituto de
Física Gleb Wataghin, UNICAMP, 13083-970, Campinas, Brazil} \affiliation{$^{3}$Los Alamos National Laboratory,
Los Alamos, New Mexico 87545, USA}

\begin{abstract}

Magnetization measurements were performed on $\mathrm{CeCoIn_{5}}$ at temperature down to 30 mK with the
magnetic field applied in three different orientations: parallel, near parallel ($\sim$ 10$^{0}$ rotated) and
perpendicular to the $ab$-plane. For these three orientations we have observed crossover features in the
torque/magnetization traces at fields just below $H_{c2}$, giving further evidence for the formation of a high
field Fulde-Ferrel-Larkin-Ovchinnikov (FFLO) superconducting state in $\mathrm{CeCoIn_{5}}$ for $H
\parallel ab$-plane and newly indicates that the FFLO state
persists for out-of-plane field orientations. Furthermore, for the
(near) parallel to $ab$-plane field configurations and $T\leq
\mathrm{50~mK}$, we have found an anomalous peak effect(APE) just
below the crossover field when the magnetic field is sweeping down
from normal to superconducting state. The dynamics of this peak
suggests the existence of a metastable phase occurring in the
vicinity of the FFLO phase and raises questions about the order (first or second) of the transition from FFLO to
the superconducting state. None of the above features were found
in a $\mathrm{Ce_{0.98}Gd_{0.02}CoIn_{5}}$ crystal.

\end{abstract}


\maketitle

The heavy-fermion (HF) $\mathrm{CeCoIn_{5}}$ is an unconventional
ambient pressure superconductor with a critical temperature $T_c$
= 2.3 K.\cite{Petrovic,Movshovich} More recently,
$\mathrm{CeCoIn_{5}}$ has attracted much interest owing to several
unusual properties which have never been displayed by any other
superconductor. For instance, a number of
studies\cite{Tayama,Radovan,Bianchi} has established that the
transition at the upper critical field $H_{c2}$ changes from
second to first-order below a temperature $T_{0}$ depending on the
field orientation, in contrast to the behavior of conventional
type-II superconductors. This fact is considered as a strong
evidence that $\mathrm{CeCoIn_{5}}$ is a Pauli limited
superconductor. In addition, there is growing experimental
evidence that $\mathrm{CeCoIn_{5}}$ is the first superconductor to
exhibit the inhomogeneous Fulde-Ferrel-Larkin-Ovchinnikov (FFLO)
superconducting (SC) state.

In the mid-1960s, theoretical studies by Fulde and
Ferrell\cite{Fulde} and Larkin and Ovchinnikov\cite{Larkin}
predicted a spatially inhomogeneous state to occur within the
mixed state of a clean superconductor in the vicinity of the upper
critical field $H_{c2}$.  The FFLO state appears when Pauli
pair-breaking overcomes the orbital effect. In the last several
years, a number of type-II superconductors, including HF and
organic compounds, have been claimed as likely candidates for the
observation of the FFLO state. However, no experimental evidence
has been accepted as an unambiguous proof for the FFLO state. With
the discovery of $\mathrm{CeCoIn_{5}}$, this issue has gained
renewed interest. Indeed, growing experimental observations
indicate the formation of the FFLO state in the high field SC
phase of $\mathrm{CeCoIn_{5}}$. For instance, heat capacity
measurements with $\mathbf{H}$ parallel to the $ab$-plane have
identified a second-order phase transition at $H_{FFLO}(T)$ within
the SC state at low temperatures and in the vicinity of
$H_{c2}$\cite{Radovan,Bianchi}. Subsequent studies, such as
thermal conductivity\cite{Capan}, ultrasound
velocity\cite{Watanabe} and nuclear magnetic
resonance\cite{Kakuyanagi} have given further evidence for the
FFLO scenario. On the other hand, results from magnetization
experiments are contradictory\cite{Radovan,Murphy,Tayama} and some
these works have claimed absence of any anomaly associated with
the FFLO state.\cite{Radovan,Tayama}

Another open issue concerns the occurrence or not of the FFLO phase for other field orientations, in
particular for $\mathbf{H}\parallel\mathbf{c}$. In fact, the number of studies for this orientation is
considerably less than for $\mathbf{H}\perp\mathbf{c}$. With the exception of the work by Bianchi et
al.\cite{Bianchi} and a more recent NMR study by Kumagai et al.\cite{Kumagai} for whose a possible FFLO phase was
identified, no evidence for the formation of such state for $\mathbf{H}\parallel\mathbf{c}$ has been given by other studies\cite{Radovan,Murphy,Tayama}.

Here, we present the results of magnetization measurements
performed on single crystals of $\mathrm{CeCoIn_{5}}$ and
$\mathrm{Ce_{0.98}Gd_{0.02}CoIn_{5}}$ at temperatures down to 30
mK. The crystals were grown by In self-flux and their phase purity
and SC transition were checked, respectively, by x-rays and
zero-field heat capacity experiments.

The measurement technique used in this work is different from refs
\cite{Tayama,Murphy} revealing new features of the magnetization
of $\mathrm{CeCoIn_{5}}$. Our experiments were carried out using a
a diaphragm force magnetometer\cite{Swanson,Bindilatti} inside a
plastic diluted refrigerator operating in a 20~T SC magnet(see
details in ref. \cite{Swanson,Bindilatti}). The magnetic force on
the sample was produced by a field gradient ($630 ~\mathrm{Oe/cm}
\leq {dH_{z}}/dz\leq 1.8~\mathrm{kOe/cm}$) superimposed on the
main magnetic field and the sample magnetic response was detected
by a capacitance technique. The contribution of the magnetic
response caused by the torque was determined by repeating the
measurement with no current in the gradient coil. From one set of
runs (with and without field gradient), we were able to extract
the component of the magnetization parallel to the magnetic field
($M_{z}$) by a simple subtraction of the contribution of the
torque from the total response for $dH_{z}/dz\neq0$. This is not
clearly the case by cantilever measurements \cite{Murphy}. Our
experimental method is similar to ref \cite{Tayama} but the force
magnetometers are different. In the present work, the movable
capacitor plate of the magnetometer is a diaphragm which gives a
stronger response to the torque than the apparatus used by Tayama
$et~al$.\cite{Sakakibara}. The data were taken for increasing and
decreasing magnetic field ($|dH/dt|\approx 35~ \mathrm{Oe/s}$)
after zero-field cooling the sample from well above
$\emph{T}_{c}$.

Figure 1 shows the capacitance response of $\mathrm{CeCoIn_{5}}$ for increasing and decreasing magnetic field at
\emph{T} = 30 mK for $\mathbf{H}\parallel\mathbf{c}$ and $\mathbf{H}\perp\mathbf{c}$ measured with and without
field gradient. When $\mathbf{H}\parallel\mathbf{c}$ (FIG.1 (a)), the response for $dH_{z}/dz\neq0$ is due only
to the magnetization ($\mathbf{M}$) parallel to the field direction ($\mathbf{M}= M_{z}$). (The capacitance
response is constant as a function of the field for $dH_{z}/dz=0$.) The traces show a sharp jump at
$H_{c2}^{\parallel}$ due to the first-order superconducting-normal state transition (FOSNT). At lower fields we
observed an hysteretic peak effect near 25 kOe which follows the same trend reported in ref\cite{Tayama}.

For $\mathbf{H}\perp\mathbf{c}$ (FIG.1 (b)), \textbf{M} is clearly
not aligned to the field direction. The response with
$dH_{z}/dz\neq0$ shows two contributions: one caused by the torque
(measured for $dH_{z}/dz$ = 0) and the other due to $M_{z}$
(displayed in FIG. 2 (b)). For both up-sweep traces (with and
without field gradient), the FOSNT manifests itself by a sharp
jump at $H_{c2}^{\perp}$. The jump at $H_{c2}^{\perp}$ in the
down-sweep traces is smaller. The data show also the existence of
a broad peak (that we will call anomalous peak effect (APE)) at
$H_{APE}$ observed only for decreasing field. This broad peak is
rapidly suppressed by thermal effects and it is absent for $T > 50
~\mathrm{mK}$. The inset presents data for the
Ce$_{0.98}$Gd$_{0.02}$CoIn$_{5}$ crystal that we discuss later.

\begin{figure}[htbp]
 \centering
     \includegraphics[width=6.5cm]{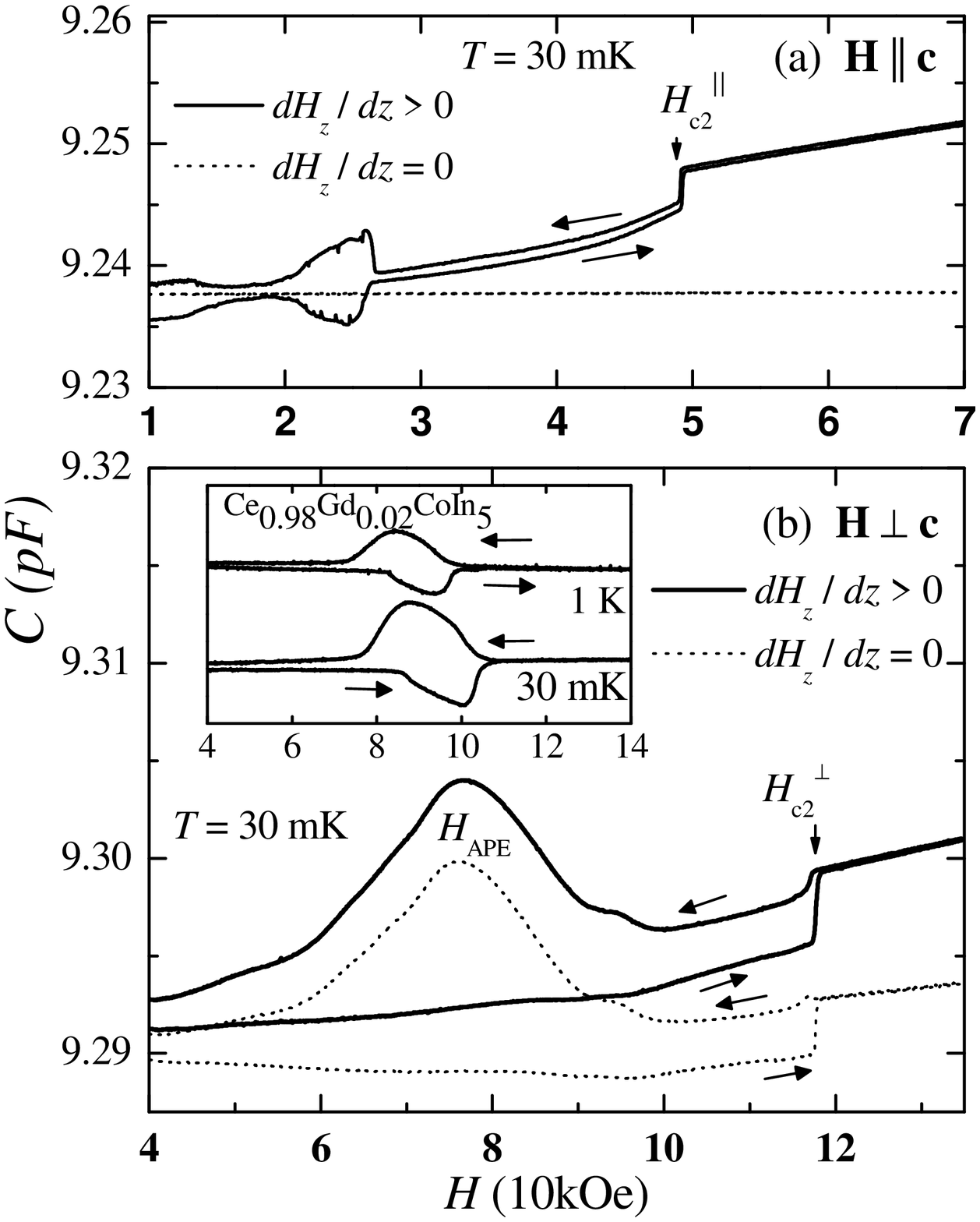}
     \vspace{-0.3cm}
    \caption{Capacitance response $C(H)$ loops for $\mathrm{CeCoIn_{5}}$ at 30 mK taken with and without field gradient
    for $\mathbf{H}\parallel\mathbf{c}$ (Fig.1 (a)) and  $\mathbf{H}\perp\mathbf{c}$ (Fig.1
    (b)). The inset shows similar set of data obtained at 30~mK and 1~K for $\mathrm{Ce_{0.98}Gd_{0.02}CoIn_{5}}$.}
    \label{Figure1}
\end{figure}

Figure 2 shows the high field part of the magnetization traces
obtained at $T$ = 30 mK for $\mathrm{CeCoIn_{5}}$. The data
presented here are the results of the average of 4 and 7
magnetization runs obtained for $\mathbf{H}\parallel\mathbf{c}$
and $\mathbf{H}\perp\mathbf{c}$ respectively. The discussed
features of the averaged trace are also observed for each
individual magnetization trace. For both orientations, the sharp
jump due to the FOSNT is clearly observed for both up and
down-sweep traces. From our data, we obtained $H_{c2}^{\parallel}$
= 49.2~kOe and $H_{c2}^{\perp}$ = 117.7~kOe with the hysteresis
width of $\mathrm{\Delta \emph{H}_{c2} \approx 0.4~kOe}$. Our
values are in very good agreement with reported data in the
literature. In addition, for both directions, we have observed a
change in the monotonic variation of the magnetization near
$H_{c2}$. This anomaly which was not yet reported in the
literature manifests itself in the derivative ($dM_{z}/dH$) trace
by a step. The observed singularity takes place at a field
$H_{\mathrm{FFLO}}$ as indicated by the vertical arrows in the
figure. For $\mathbf{H}\parallel\mathbf{c}$, $H_{\mathrm{FFLO}}$
is found to be nearly independent on $T$. On the other hand, for
$\mathbf{H}\parallel\mathbf{c}$ the $H_{\mathrm{FFLO}}$ is clearly
shifted to higher fields for increasing temperature.

\begin{figure}[htbp]
    \centering
    \vspace{-0.2cm}
        \includegraphics[width=7.0cm]{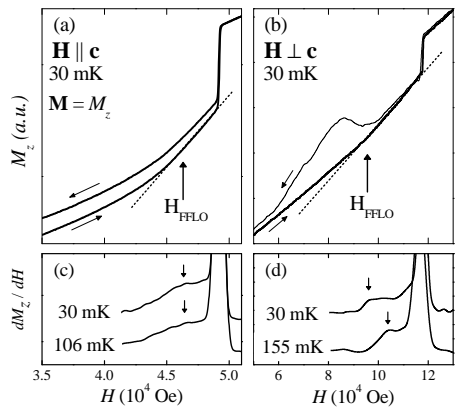}
          \vspace{-0.2cm}
        \caption{High-field magnetization of $\mathrm{CeCoIn}_{5}$ for $\mathbf{H}\parallel~\mathbf{c}$ and
           $\mathbf{H}~\perp~\mathbf{c}$ at $T = 30\;\mathrm{ mK}$. The dotted lines emphasize the change in the monotonic
           variation of ${M_{z}(H)}$ near $H_{c2}$. For $\mathbf{H}~\perp~\mathbf{c}$, the APE is seen in the down-sweep trace. The
             respective $dM_{z}/dH$ traces are shown in the Fig. 2 (c)
           and (d). for two temperatures.
           The vertical arrows indicate $H_{\mathrm{FFLO}}$ in both magnetization and derivative traces.
           The other arrows indicate the direction of the field sweep.}
    \label{Figure2}
\vspace{-0.2cm}
\end{figure}

\begin{figure}[htbp]
    \centering
    \includegraphics[width=8cm]{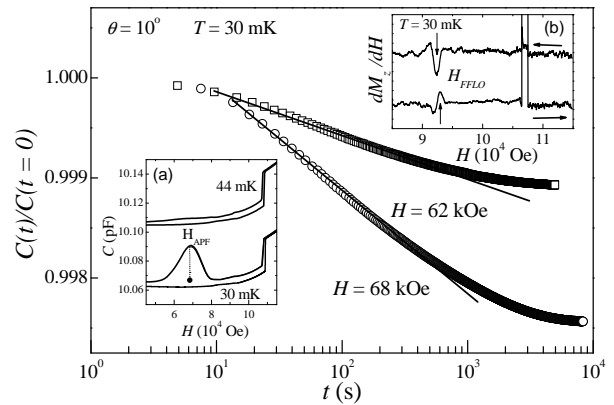}
    \vspace{-0.2cm}
    \caption{Results of $C(t)$ for $\mathrm{CeCoIn}_{5}$ obtained at $T$ = 30 mK for two values of magnetic
    field applied $\sim$ 10° from the $ab$-planes. (a) $C(H)$ at two temperatures. The dashed line at the field $H_{\mathrm{APE}}$
    denotes the time decay $C(t)$. (b) Up and down traces of $dM_{z}/dH$ as a function of field at 30~mK. The arrows indicate $H_{\mathrm{FFLO}}$ for the two sweep directions.}
    \label{Figure3}
    \end{figure}

To further explore the existence of the FFLO phase and the nature
of the broad peak at $H_{\mathrm{APE}}$ we have performed
additional experiments. Firstly, we repeat the measurements shown
in Fig.1 for the magnetic field slightly rotated from the basal
plane of the crystal. Figure~\ref{Figure3} shows in the inset (a)
the results for $dH_{z}/dz$ = 0 up/down sweep traces at
$T$~=~30~mK and $T$~=~44~mK. For $T$~=~30~mK, the FOSNT occurs at
$H_{c2}$ = 109.6 kOe which is consistent with a misalignment of
roughly 10$^{\circ}$ with respect to the $ab$-plane \cite{Murphy}.
The $H_{\mathrm{FFLO}}$-anomaly (see inset (b) in FIG.~3) was also
observed in the $M_{z}(H)$ traces at slightly lower field than
that for $\mathbf{H}\perp\mathbf{c}$ and the position of the broad
peak $H_\mathrm{{APE}}$ is shifted to lower field as well.
Secondly, we performed time relaxation measurements at selected
values of the field in the region of the broad peak observed for
$\mathbf{H}\parallel \mathbf{ab}$-plane and for
$\theta~\sim~10^{\circ}$. Before each measurement, the field was
sweeping up to 140 kOe, subsequently sweeping down to the target
value of the field and then the signal was measured as a function
of time (see inset (a) in Fig.\ref{Figure3}). The sweep rate was
identical in all these measurements. The main panel of
Fig.\ref{Figure3} shows the time dependence of $C(t)$ $\propto$
$\tau(t)$ $\propto$ $M(t)$ for two fields in the region of the
peak for the $\theta~\sim~10^{\circ}$ orientation. We concentrate
on this particular orientation where the broad peak
$H_\mathrm{{APE}}$ is even more pronounced than for
$\mathbf{H}\perp\mathbf{c}$ (see Figs. 1 and 3) and the FOSNT is
clearly observed for both up and down-sweep traces. But similar
time-dependence was also found in the peak region for
$\mathbf{H}\perp\mathbf{c}$. From the results displayed in
Fig.\ref{Figure3}, we conclude that the broad peak at $H_{APE}$ in
down-sweep traces for the $\mathbf{H}\parallel \mathbf{ab}$-plane
and $\theta~\sim~10^{\circ}$ geometries is related to a metastable
state.

\begin{figure}[htbp]
\vspace{-0.8cm}
    \centering
             \includegraphics[width=5cm]{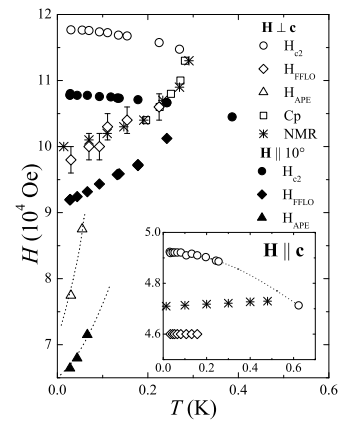}
             \vspace{-0.4cm}
                   \caption{$H-T$ phase diagram of CeCoIn$_{5}$ for $\textbf{H}~\perp~\textbf{c}$,
                   $\textbf{H}\parallel 10^{\circ}$ and $\textbf{H}~\parallel~\textbf{c}$.
                   The graph includes results from heat capacity\cite{Bianchi} and NMR\cite{Kakuyanagi,Kumagai}.Dotted lines are guides
                   to the eyes.}
    \label{Figure4}
\vspace{-0.6cm}
\end{figure}

By mapping the $T$-dependence of $H_{c2}$, $H_{\mathrm{FFLO}}$ and
$H_{\mathrm{APE}}$, we have constructed the high-field
low-temperature phase diagram for $\mathrm{CeCoIn}_{5}$ displayed
in FIG.\ref{Figure4}. Results from heat capacity\cite{Bianchi} and
NMR\cite{Kakuyanagi,Kumagai} experiments are also shown for
completeness. The evolution of the $H_{\mathrm{FFLO}}$ line
determined from our magnetic measurements is in good agreement
with the line identified by distinct techniques. Furthermore,
although it takes place only at $T~<~50$ mK, the
$H_{\mathrm{APE}}$ line has the same behavior as the FFLO line.

In what follows we speculate on the origin of the APE found below
the FFLO line for $H$ (near)parallel to the $ab$-planes. In
general, the magnetic relaxation (time dependence of the
irreversible magnetization) in type-II superconductors originates
from the vortex motion driven by the gradient in the vortex
density $\nabla{n_v}$, in presence of the vortex pinning
(so-called vortex creep regime). For our studies in the (nearly)
parallel to the $ab$-planes field configurations the torque is
proportional to the $c$-axis component of the magnetization
M$_{c}$(t). Then, in the vortex creep regime, the time dependent
signal $C(t)$ $\propto$ $\tau(t)$ $\propto$ $M_c(t)$. Because
CeCoIn$_{5}$ is anisotropic (layered) superconductor, the
appearance of vortex kinks along the $c$-axis and their
interaction with vortex systems parallel to $ab$-planes should
also be taken into account; this would provide the information on
pinning of both in- and out-of-plane vortex
systems\cite{Brongersma}. In principle, our observation of the
logarithmic time relaxation (see Eq. 1 below) of the measured
signal $C(t)~\sim~M(t)$ (FIG. 3) in the peak region
$55~kOe~\leq~H~\leq~80~kOe$ for the $\theta$ $\sim$ 10$^{\circ}$
orientation, agrees with the above scenario, and suggests that the
vortex pinning efficiency is enhanced in this field interval.
However, in contrast to the more conventional peak effect behavior
observed for $\mathbf{H}\parallel~\mathbf{c}$ (see Fig. 1(a) and
Refs.\cite{Tayama}), the APE measured for
$\mathbf{H}\perp\mathbf{c}$ and $\mathbf{H}\parallel10^{\circ}$
(see Fig. 1(a) and Fig. 3), takes place only under field
decreasing at very low temperatures $T~\leq~50~mK$. Besides, there
is an order of magnitude difference in the effective activation
energy values $U_e(H,T)$ as obtained from the Kim-Anderson
equation \cite{Tinkham}:
\begin{equation}
M(t)=M_0(1-\frac{k_BT}{U_e})\ln(\frac{t}{t_0})
\end{equation}

Taking characteristic value $\ln(t/t_0)=15$ \cite{Tinkham}, we find for the two field geometries in the peak
effect regions $U_e(H\perp~c,H_{peak}=6.84~T)\sim10^3$~K and $U_e(H\parallel~c,H_{peak}=2.46~T)\sim~10^4$~K.
Taking into account the relatively small anisotropy $\gamma=(m_c/m_{ab})^{1/2}\sim2$ (\cite{Xiao} and refs.
therein) as well as nearly the same reduced fields $(H_{peak}/H_{c2})_{||c}~=~0.5$ and
$(H_{peak}/H_{c2})_{\perp~c}~=~0.58$, our results suggest essentially different mechanisms responsible for the
peak effects measured in $\mathbf{H}\parallel~\mathbf{c}$ and $\mathbf{H}\perp\mathbf{c}$ field configurations.

In attempts to shed light on origin of the APE, we note that irreversible under increasing/decreasing field
state, characterized by a glassy-like time relaxation, has been also reported for others spin-paramagnetically
limited superconductors such as Al \cite{Wu,Wenhao,Yu,Butko} and Be \cite{Adams} films. Interestingly, the low
temperature/high magnetic field portion of the $H-T$ plane (see Fig. 4), resembles very much the metastable $H-T$
phases identified in Al and Be films, where SC and normal (N) states coexist.\cite{Wu,Wenhao,Yu,Butko,Adams}

All these may indicate that the pronounced magnetization time
relaxation occurring in $\mathrm{CeCoIn}_{5}$ for $\mathbf{H}$
nearly parallel/parallel to the $ab$-planes may be governed by
dynamics of coexisting SC and N domains, or possibly BCS-like
superconducting (vortex state (VS) in our case) and FFLO states.
Such possibility has been considered in the context of
superfluidity of atomic fermion gases.\cite{Sheehy} Based on the
experimental data of Fig. 1(b), the coexistence of N and SC
domains cannot be ruled out for $\mathbf{H}\perp\mathbf{c}$
geometry, indeed. However, the results presented in Fig. 3 for the
inclined field indicate that the APE takes place well outside of
the field hysteresis region associated with the FOSNT.  Thus,
VS-FFLO domain coexistence looks a more suitable scenario. The
domain boundaries can act as additional pinning centers for
vortices leading to the APE. If our interpretation is correct, the
FFLO line may represent in fact a first order transition, as
theoretically predicted\cite{Gruenberg}, giving rise to a
metastable state with VS-FFLO domain coexistence in its vicinity.
Whereas specific heat measurements\cite{Bianchi} suggest the
second order VS-FFLO transition, quenched disorder can mask the
first order nature of the transition\cite{Imry}. In fact in the
closest-to-equilibrium magnetization data which is the up-sweep
trace curve for $\theta~\sim~10^{\circ}$ orientation (see inset
(b) Fig. 3), the  peak-like (rather than step-like (Fig.~2))
character of the $dM_{z}/dH$ anomaly at $H_{\mathrm{FFLO}}$
suggest a first order FFLO-VS transition. Finally, if all the
features discussed in the text are indeed associated to the FFLO
state, they should disappear as sample is driven away from the
clean limit. To verify this idea we have performed similar
experiments for $\mathrm{Ce_{0.98}Gd_{0.02}CoIn_{5}}$ single
crystal (inset of Fig. 1). According to our data, the substitution
of Gd suppresses the FOSNT at $H_{c2}^{\perp}$. Simultaneously,
the FFLO state ($\mathbf{H}\perp\mathbf{c}$) vanishes, the peak
effect becomes conventional and remains to higher-$T$, and
magnetic relaxation is invisible for the same temperature and time
window (not shown) as measured for pure CeCoIn$_{5}$.

In conclusion, by torque/magnetization measurements we have found
crossing over anomalies at the vicinity of $H_{c2}$ that are in
agreement with $H_{FFLO}(T)$ line determined by
others\cite{Radovan,Bianchi,Capan,Watanabe} for
$\mathbf{H}\perp\mathbf{c}$. Besides, our data indicates that the
FFLO state persists for out-of-plane field orientations and
reveals an APE, possibly associated with a metastable new phase
below the FFLO line. This results suggest that the transition at
the FFLO line may be indeed weakly first order. We hope that the
new results reported in this work will stimulate further
experimental and theoretical studies in the vicinity of FFLO phase
in CeCoIn$_{5}$. We thank A. D. Bianchi for fruitful discussions
and Fapesp-SP, CNPq-Brazil and US DOE for supporting this work.

\end{document}